\documentclass[twocolumn, prl]{revtex4-1}
\usepackage{amsmath,amssymb,amsthm,epsfig,graphics,graphicx,hyperref}
\usepackage{verbatim}
\usepackage{color}

\newtheorem{theorem}{Theorem}
\newtheorem{lemma}{Lemma}

\theoremstyle{definition}

\newtheorem{proposition}{Proposition}

\newcommand{\bra}[1]{\langle #1|}
\newcommand{\ket}[1]{|#1\rangle}

\newcommand{\ip}[2]{\langle #1|#2\rangle}
\newcommand{\op}[2]{|#1\rangle \langle #2|}

\def\0{{\mathbf{0}}}
\def\1{{\mathbf{1}}}
\def\2{{\mathbf{2}}}
\def\3{{\mathbf{3}}}
\def\4{{\mathbf{4}}}
\def\5{{\mathbf{5}}}
\def\6{{\mathbf{6}}}
\def\7{{\mathbf{7}}}
\def\8{{\mathbf{8}}}
\def\9{{\mathbf{9}}}

\begin{document}

\title{A Return to the Optimal Detection of Quantum Information}

\author{Eric Chitambar $^1$}
\email{echitamb@siu.edu}
\author{Min-Hsiu Hsieh $^2$}
\email{minhsiuh@gmail.com}

\affiliation{$^1$ Department of Physics and Astronomy{\mbox ,} Southern Illinois University, 
Carbondale, Illinois 62901, USA\\
$^2$ Centre for Quantum Computation \& Intelligent Systems (QCIS), Faculty of Engineering and Information Technology (FEIT), University of Technology Sydney (UTS), NSW 2007, Australia}

\date{\today}

\begin{abstract}

In 1991, Asher Peres and William Wootters wrote a seminal paper on the nonlocal processing of quantum information [\textit{Phys. Rev. Lett.} \textbf{66} 1119 (1991)].  We return to their classic problem and solve it in various contexts. Specifically, for discriminating the ``double trine'' ensemble with minimum error, we prove that global operations are more powerful than local operations with classical communication (LOCC). Even stronger, there exists a finite
gap between the optimal LOCC probability and that obtainable by separable
operations (SEP).   Additionally we prove that a two-way, adaptive LOCC strategy can always beat a one-way protocol.  Our results provide the first known instance of ``nonlocality without entanglement'' in two qubit pure states.  
\end{abstract}

\maketitle

One physical restriction that naturally emerges in quantum communication scenarios is nonlocality.  Here, two or more parties share some multi-part quantum system, but their subsystems remain localized with no ``global'' quantum interactions occurring between them.  Instead, the system is manipulated through local quantum operations and classical communication (LOCC) performed by the parties.

Asher Peres and William Wootters were the first to introduce the LOCC paradigm and study it as a restricted class of operations in their seminal work \cite{Peres-1991a}.  To gain insight into how the LOCC restriction affects information processing, they considered a seemingly simple problem.  Suppose that Alice and Bob each possess a qubit, and with equal probability, their joint system is prepared in one of the states belonging to the set $\{\ket{D_i}=\ket{s_i}\otimes\ket{s_i}\}_{i=0}^2$, where $\ket{s_i}=U^i\ket{0}$ and $U=exp(-\tfrac{i\pi}{3}\sigma_y)$.  This highly symmetric ensemble is known as the ``double trine,'' and we note that lying orthogonal to all three states is the singlet $\ket{\Psi^-}=\sqrt{1/2}(\ket{01}-\ket{10})$.  

Alice and Bob's goal is to identify which double trine element was prepared only by performing LOCC.  Like any quantum operation used for state identification, Alice and Bob's collective action can be described by some \textit{positive-operator valued measure} (POVM).  While the non-orthogonality of the states prohibits the duo from perfectly identifying their state, there are various ways to measure how well they can do.  Peres and Wootters chose the notoriously difficult measure of accessible information \cite{Holevo-1973a, *Nielsen-2000a},  
but their paper raises the following two general conjectures concerning the double trine ensemble, which can apply to any measure of distinguishability:
\begin{enumerate}
\item [C1:]  LOCC is strictly sub-optimal compared to global operations,
\item [C2:]  The optimal LOCC protocol involves two-way communication and adaptive measurements.
\end{enumerate}
The set of global POVMs will be denoted by GLOBAL, and C1 can be symbolized by GLOBAL $>$ LOCC.  A two-way LOCC protocol with adaptive measurement refers to \textit{at least} three rounds of measurement, Alice $\to$ Bob $\to$ Alice, where the choice of measurement in each round depends on the outcome of the other party's measurement in the previous round.  
We symbolize C2 as LOCC $>$ LOCC$_\rightarrow$.  In Ref. \cite{Peres-1991a} Peres and Wootters obtained numerical data to support both C1 and C2, but these conjectures have never been proven for the double trine.

Before we present our contribution to the problem, we would like to briefly highlight the legacy of the Peres-Wootters paper.  Perhaps most notably is that it subsequently led to the discovery of quantum teleportation \cite{Bennett-1993a}.  Other celebrated phenomena can also directly trace their roots to Ref. \cite{Peres-1991a} such as so-called nonlocality without entanglement \cite{Bennett-1999a} and quantum data hiding \cite{Terhal-2001a, *DiVincenzo-2002a}.  More generally, Ref. \cite{Peres-1991a} paved the way for future research into LOCC and its fundamental connection to quantum entanglement \cite{Horodecki-2009a}.


We finally note that in a return to Ref. \cite{Peres-1991a} of his own, Wootters constructed a \textit{separable} POVM that obtains the same information as the best known global measurement \cite{Wootters-2005a}.  A POVM $\{\Pi_i\}$ belongs to the class of separable operations (SEP) if each POVM element can be decomposed as a tensor product $\Pi_i=A_i\otimes B_i$ over the two systems.  SEP is an important class of operations since every LOCC operation belongs to SEP \cite{Bennett-1999a}.  


In this paper, we prove that conjectures C1 and C2 are indeed true when distinguishability success is measured by the \textit{minimum error probability}, which is defined as follows.  For an ensemble $\mathcal{E}=\{\ket{\psi_i},p_i\}_{i=1}^k$, the error probability associated with some identification POVM $\{\Pi_i\}_{i=1}^k$ is given by $1-\sum_{i=1}^kp_i\bra{\psi_i}\Pi_i\ket{\psi_i}$.  Then the minimum error probability of distinguishing $\mathcal{E}$ with respect to a class of operations $\mathcal{S}$ (such as LOCC, SEP, GLOBAL, etc.) is given by the \textit{infimum} of error probabilities taken over all POVMs that can be generated by $\mathcal{S}$.  
Note that we can replace ``infimum'' by ``minimum'' only if $\mathcal{S}$ is a compact set of operations.  While GLOBAL, SEP and LOCC$_\rightarrow$ all have this property, LOCC does not \cite{Chitambar-2012a, Chitambar-2012c}.  Hence, to properly discuss the LOCC minimum error, we must consider the class of so-called \textit{asymptotic LOCC}, which is LOCC plus all its limit operations \cite{Chitambar-2012c}.  We will prove C1 with respect to this more general class of operations. 

\textit{A. Global and Separable Operations}:  The double trine ensemble has a group-covariant structure which greatly simplifies the analysis.  In fact, Ban \textit{et al.} have proven that the so-called ``Pretty Good Measurement'' (PGM) \footnote{Recall that the ``Pretty Good Measurement'' for an ensemble $\{\ket{\phi_i},p_i\}_{i=1}^k$ is the POVM with elements \unexpanded{$p_i\rho^{-1/2}\op{\phi_i}{\phi_i}\rho^{-1/2}$, where $\rho=\sum_{i=1}^kp_i\op{\phi_i}{\phi_i}$} \cite{Hausladen-1994a}.} is indeed an optimal global POVM for discriminating ensembles with such symmetries \cite{Ban-1997a}.  For the double trine, the PGM consists of simply projecting onto the orthonormal basis $\{\ket{\Psi^-},U^i\otimes U^i\ket{F_i}\}_{i=0}^2$, where
\begin{equation}
\label{Eq:GlobalPOVM}
\ket{F_i}\propto U^i\otimes U^i[(\sqrt{2}+1)\ket{00}-(\sqrt{2}-1)\ket{11}].
\end{equation}
The corresponding error probability is 
\begin{equation}
\label{Eq:GlobalProb}
1/2-\sqrt{2}/3\approx 2.86\times 10^{-2}.
\end{equation}
To show that SEP can also obtain this probability, we explicitly construct a separable POVM.  The idea is to mix a sufficient amount of the singlet state with each of the PGM POVM elements so to obtain separability (a similar strategy was employed in Ref. \cite{Wootters-2005a}).  The resulting POVM is $\{\op{\tilde{F}_i}{\tilde{F}_i}\}_{i=0}^2$ with $\op{\tilde{F}_i}{\tilde{F}_i}=\op{F_i}{F_i}+1/3\op{\Psi^-}{\Psi^-}$.  It is fairly straightforward to compute that $\tilde{F}_0=1/2(\op{\varphi_+}{\varphi_+}+\op{\varphi_-}{\varphi_-})$, where $\ket{\varphi_{\pm}}=\ket{F_0}\pm\sqrt{1/3}\ket{\Psi^-}$ is a product state.  This suffices to prove separability of the POVM.   

\textit{B. LOCC and Asymptotic LOCC}:  
Let us begin with a clear description of asymptotic LOCC discrimination.  In general, a sequence of POVMs $\mathcal{P}^{(n)}:=\{\Pi^{(n)}_i\}_{i=1}^k$ \textit{asymptotically attains} an error probability $P$ on ensemble $\{\ket{\psi_i},p_i\}_{i=1}^k$ if for every $\epsilon>0$ we have $P+\epsilon>1-\sum_{i=1}^kp_i\bra{\psi_i}\Pi_i^{(n)}\ket{\psi_i}$ for sufficiently large $n$.  If each POVM in the sequence $\mathcal{P}^{(n)}$ can be generated by LOCC, then $P$ is achievable by asymptotic LOCC.  

It is known that for an ensemble of linearly independent pure states, the global POVM attaining minimum error consists of orthonormal, rank one projectors \cite{Yuen-1975a} (see also \cite{Mochon-2006a}).  We strengthen this result and extend it to the asymptotic setting.
\begin{theorem}
\label{Thm1}
Let $\mathcal{E}=\{\ket{\psi_i},p_i\}_{i=1}^k$ be an ensemble of linearly independent states spanning some space $S$.  Suppose that $P_{opt}$ is the global minimum error probability of $\mathcal{E}$.  Then there exists a unique orthonormal basis $\{\ket{\phi_i}\}_{i=1}^k$ of $S$ such that: (a) A POVM attains an error probability $P_{opt}$ on $\mathcal{E}$ if and only if it can also distinguish the $\{\ket{\phi_i}\}_{i=1}^k$ with no error, and (b) A sequence of POVMs asymptotically attains an error probability $P_{opt}$ on $\mathcal{E}$ if and only if it contains a subsequence that can asymptotically distinguish the $\{\ket{\phi_i}\}_{i=1}^k$ with no error.
\end{theorem}
The proof is given in the Appendix.  Theorem \ref{Thm1} essentially reduces optimal distinguishability of non-orthogonal linearly independent ensembles to perfect discrimination of orthogonal ensembles.  Applying part (a) to the double trine ensemble, if an LOCC POVM could attain the error probability of Eq. \eqref{Eq:GlobalProb}, then it can also perfectly distinguish the states $\ket{F_i}$ given by \eqref{Eq:GlobalPOVM}.  However, these are three entangled states which, by a result of Walgate and Hardy, means they cannot be distinguished perfectly by LOCC \cite{Walgate-2002a}.  Therefore, the global minimum error probability is unattainable by LOCC. 

But is the probability attainable by asymptotic LOCC?  If it is, then part (b) of Theorem \ref{Thm1} likewise implies that the $\ket{F_i}$ must be perfectly distinguishable by asymptotic LOCC.  While Ref. \cite{Walgate-2002a} provides simple criteria for deciding perfect LOCC distinguishability of two qubit ensembles, no analogous criteria exists for asymptotic LOCC.  The only general result for asymptotic discrimination has been recently obtained by Kleinmann \textit{et al.} \cite{Kleinmann-2011a}.  Here we cite their result in its strongest form, adapted specifically for the problem at hand.
\begin{proposition}[\cite{Kleinmann-2011a}]
\label{Prop1}
If the states $\{\ket{F_i}\}_{i=0}^2$ can be perfectly distinguished by asymptotic LOCC, then for all $\chi\in[1/3,1]$ there is a product operator $E\geq 0$ such that (i) $\sum_{i=0}^2\bra{F_i}E\ket{F_i}=1$, (ii) $\bra{F_0}E\ket{F_0}=\chi$, and (iii) the normalized states $\ket{F'_i}:=\tfrac{1}{\sqrt{\bra{F_i}E\ket{F_i}}}E^{1/2}\ket{F_i}$ are perfectly distinguishable by separable operations.
\end{proposition} 
\noindent In the appendix we prove that these three conditions cannot be simultaneously satisfied; therefore, GLOBAL $>$ LOCC for minimum error discrimination.  Here, we provide a little intuition into why Proposition \ref{Prop1} must be true.  For every LOCC protocol that correctly identifies the given state with probability $1-\epsilon$, we can think of the success probability as smoothly evolving from complete randomness ($\chi=1/3$) to its final average value ($\chi=1-\epsilon$).  Then for each $\chi\in(1/3,1-\epsilon)$, the protocol can be halted after some sequence of measurement outcomes (collectively represented by the product operator $E$) such that given these outcomes: (1) there is one state that can be identified with probability $\chi$ (which by symmetry we can assume is $\ket{F_0}$), and (2) the transformed ensemble can be discriminated by a separable POVM with success probability no less than $1-\epsilon$.  By compactness of SEP, we let $\epsilon\to 0$ and replace (2) by the condition that a separable POVM perfectly distinguishes the post-halted ensemble.

\textit{C. LOCC $>$ LOCC$_\rightarrow$:}  
We will now compute the minimum one-way error probability for the double trine, and then describe an explicit two-way protocol with a smaller error probability.  In the one-way task, Alice makes a measurement and communicates her result to Bob.  Without loss of generality, we fine-grain Alice's measurement so that each POVM element is rank one $\op{\eta}{\eta}$, with $\ket{\eta}=r\cos\theta\ket{0}+re^{i\phi}\sin\theta\ket{1}$.  Given outcome $\eta$, Bob's task is to optimally discriminate the ensemble $\{\ket{s_i}\}_{i=0}^2$, but now with an updated distribution $\{p_{i}\}_{i=0}^2$ given by
\begin{align}
\label{Eq:probsBob}
p_{k}&=\tfrac{|\ip{\eta}{s_k}|^2}{3P(\eta)}=\tfrac{2}{3}|\cos\tfrac{2\pi k}{3}\cos\theta+e^{i\phi}\sin\tfrac{2\pi k}{3}\sin\theta|^2.
\end{align}  
Here, $P(\eta)=\tfrac{1}{3}\sum_{i=0}^2|\ip{\eta}{s_i}|^2$, and we've used the covariance $\frac{1}{3}\sum_{i=0}^2\op{s_i}{s_i}=\mathbb{I}/2$.  Additionally, we can assume that $p_{0}\geq p_{1},p_{2}$, since if $\ket{\eta}$ fails to generate a distribution with this property, by the symmetry we can always rotate $\ket{\eta}$ such that $p_{0}$ is indeed the maximum post-measurement probability.  This means we can only restrict attention to $-\pi/6\leq \theta\leq \pi/6$. 

Next, we observe that Bob's task of distinguishing the ensemble $\{\ket{s_i},p_{i}\}_{i=0}^2$ is no easier than distinguishing between the two weighted states $\rho=p_0\op{s_0}{s_0}$ and $\sigma=p_{1}\op{s_1}{s_1}+p_{2}\op{s_2}{s_2}$.  Indeed, any protocol distinguishing the three $\ket{s_i}$ can always be converted into a protocol for distinguishing $\rho$ and $\sigma$ by simply coarse-graining over all outcomes corresponding to $\ket{s_2}$ and $\ket{s_3}$.  The minimum error probability in distinguishing $\rho$ and $\sigma$ is readily found to be (see Appendix): 
\begin{equation}
\label{Eq:mixedminerr}
\tfrac{1}{2}-\tfrac{1}{2}\sqrt{1-3p_{1}p_{2}-p_{0} p_{1}-p_{0} p_{2}},
\end{equation}
which simplifies to $\tfrac{1}{2}-\tfrac{1}{24}[75 + 32 \cos(2\theta) - 7 \cos (4\theta)+18\cos(2\phi)\sin^2(2\theta)]^{1/2}$.  In the interval  $-\pi/6\leq \theta< \pi/6$, a minimum is obtained at $\theta=-\pi/6$ and $\phi=0$.  This corresponds to $p_0=p_1=1/2$ and $p_2=0$ with an error probability of $1/2-\sqrt{3}/4$.  Now, this probability lower bounds the error probability along each branch of Alice's measurement, and therefore it places a lower bound on any one-way LOCC measurement scheme.  In fact, this lower bound turns out to be tight.  When Alice performs the POVM $\{\frac{2}{3}(\mathbb{I}-\op{s_i}{s_i})\}_{i=0}^2$ outcome $i$ will eliminate $\ket{s_i}\otimes\ket{s_i}$ but leave the other two states with an equal post-measurement probability.  Thus, in each branch we obtain the error probability  $1/2-\sqrt{3}/4\approx 6.70\times 10^{-2}$, and this provides the minimum one-way error probability.

\begin{figure}[h]
\includegraphics[scale=0.6]{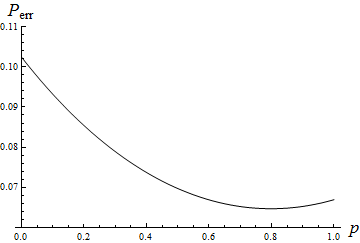}
\caption{\label{Fig:LOCC2}
The error probability $P_{err}$ using the above protocol as a function of Alice's measurement strength $p$.  The point $p=1$ is the one-way minimum error probability. } 
\end{figure}

If we allow feedback from Bob, there exists better measurement strategies.  The following protocol generalizes the optimal one-way scheme just described. (Round I) Alice performs the measurement with Kraus operators given by $\{A_i\}_{i=0}^2$ with \[A_i=\sqrt{1/3(1-p)}\op{s_i}{s_i}+\sqrt{1/3(1+p)}\op{s_i^\perp}{s_i^\perp}.\]  
Here $\ket{s_i^\perp}$ is the state orthogonal to $\ket{s_i}$ (explicitly $\ket{s_i^\perp}=U^i\ket{1}$).  Note that this is the square-root of the POVM given by Peres and Wootters \cite{Peres-1991a}.  Without loss of generality, we suppose that Alice obtains outcome ``$0$'' and communicates the result to Bob.  Her (normalized) post-measurement states are $\ket{s'_0}=\ket{0}$, $\ket{s_1'}=[2(2+p)]^{-1/2}(\sqrt{1-p}\ket{0}-\sqrt{3(1+p)}\ket{1})$, and $\ket{s_2'}=[2(2+p)]^{-1/2}(\sqrt{1-p}\ket{0}+\sqrt{3(1+p)}\ket{1})$.  (Round II) From Bob's perspective, he is still dealing with the original states $\ket{s_i}$, but now their prior probabilities have changed to $P_{i|A_0}=P_{A_0|i}$.  He now proceeds as if Alice had completely eliminated the state $\ket{s_{0}}$ (i.e. if she had chosen $p=1$ as the strength of her measurement).  Specifically, he projects onto the eigenbasis of $\op{s_1}{s_1}-\op{s_2}{s_2}$ which are the states $\ket{\pm}=\sqrt{1/2}(\ket{0}\pm\ket{1})$.   A ``$+$'' outcome is associated with $\ket{s_1}$ and a ``$-$'' outcome is associated with $\ket{s_2}$; this is the optimal measurement for distinguishing between two pure states \cite{Helstrom-1976a}.  By the symmetry of the states, it is sufficient to only consider the ``$+$'' outcome, which he communicates to Alice.  The conditional probabilities are $P_{A_0B_+|0}=(1-p)/6$, $P_{A_0B_+|1}=1/24 (2 + \sqrt{3}) (2 + p)$, and $P_{A_0B_+|2}=1/24 (2 - \sqrt{3}) (2 + p)$.  These can be inverted to give $P_{i|A_0B_+}=2P_{A_0B_+|i}$.  (Round III)  At this point, Alice still has three distinct states $\ket{s_0'}$, $\ket{s_1'}$ and $\ket{s_2'}$.  Here, $\ket{s_1'}$ will have the greatest probability while $\ket{s_0'}$ will have the smallest when $p$ is close to $1$.  Alice then ignores $\ket{s_2'}$ and performs optimal discrimination between just $\ket{s_{0}'}$ and $\ket{s_{1}'}$.  Letting $Q=P_{0|A_0B_+}+P_{1|A_0B_+}$, the minimum error probability is given by the well-known \textit{Helstrom bound} \cite{Helstrom-1976a} with normalized probabilities: 
\[P_{err}^{(A_0B_+)}=1-\frac{Q}{2}(1+\sqrt{1-4\frac{P_{0|A_0B_+}P_{1|A_0B_+}}{Q^2}|\ip{s_0'}{s_1'}|^2}).\]
By symmetry, each sequence of outcomes $(A_i,B_\mu)$ - with $i\in\{0,1,2\}$, $\mu\in\{+,-\}$ - occurs with the same probability.  Hence, the total error probability across all branches is given by $P_{err}=6P_{err}^{(A_0B_+)}$.  The plot is given in Fig. \ref{Fig:LOCC2}.  It obtains a minimum of approximately $6.47\times 10^{-2}$, which is smaller than the one-way optimal of $1/2-\sqrt{3}/4\approx 6.70\times 10^{-2}$.  The one-way optimal probability is obtained at the point $p=1$.

\textit{Discussion and Conclusions:} Our results for minimum error discrimination of the double trine ensemble can be summarized as:
\[\text{GLOBAL} = \text{SEP} > \text{LOCC} > \text{LOCC}_\rightarrow.\]
We thus put substantial closure to a problem first posed over 20 years ago.  A primary motivation for studying this problem is to better understand the limitations of processing quantum information by LOCC.  Our results complement a series of recent results in this direction \cite{Kleinmann-2011a, Childs-2012a, Chitambar-2012c}.  In particular is Ref. \cite{Kleinmann-2011a} which provides a necessary condition for perfect discrimination by asymptotic LOCC discrimination (Prop. \ref{Prop1} above).  Theorem \ref{Thm1} of our paper largely extends this result as we reduce asymptotic minimum error discrimination of linearly independent states to asymptotic perfect discrimination. 

Our proofs of C1 and C2 are the first of its kind for two qubit ensembles, and we contrast it with previous work on the subject.  C1 was first shown by Massar and Popescu for two qubits randomly polarized in the same direction \cite{Massar-1995a}.  However, a different distinguishability measure was used and the asymptotic case was not considered.  Later, Koashi \textit{et al.} showed an asymptotic form of C1 for two qubit \textit{mixed} states with respect to the different task of ``unambiguous discrimination'' \cite{Koashi-2007a} (the same can also be shown for the double trine ensemble \cite{Chitambar-2013a}).  Finally, C2 has been observed by Owari and Hayashi on mixed states and only for a special sort of distinguishability measure \cite{Owari-2008a}.  Our work is distinct from all previous results in that it deals with pure states and minimum error probability, a highly natural measure of distinguishability.  The fact that we consider pure ensembles with three states is significant since it is well-known that any \textit{two} pure states can be distinguished optimally via LOCC (i.e. LOCC $=$ GLOBAL) \cite{Walgate-2000a, Virmani-2001a}.  Thus, with the double trine being a real, symmetric, and pure ensemble of two qubits, we have identified the simplest type of ensemble in which LOCC $\not=$ GLOBAL for state discrimination.  

Even more, since the double trine ensemble consists of product states (i.e. no entanglement), we have shown that ``nonlocality without entanglement'' can exist in even the simplest types of ensembles with more than two states.  This distinction is further sharpened by considering that LOCC $\not=$ SEP for the optimal discrimination of the double trine.  Separable operations are interesting since, like LOCC operations, they lack the ability to create entanglement.  Nevertheless, SEP evidently possesses some nonlocal power as it can outperform LOCC in discriminating the double trine.  Thus, entanglement and nonlocality can truly be regarded as two distinct resources, even when dealing with two qubit pure states.



\begin{acknowledgments}
We would like to thank Runyao Duan, Debbie Leung, and Laura Man\v{c}inska for helpful discussions on the topic of LOCC distinguishability.
\end{acknowledgments}

\appendix
\section{Appendix}
\subsection{Proof of Theorem \ref{Thm1}}
(a)  We first recall a few general facts about minimum error discrimination.  A POVM $\{\Pi_i\}_{i=1}^k$ is optimal on $\mathcal{E}$ if and only if $\Lambda\geq p_j\op{\psi_j}{\psi_j}$ for all $\ket{\psi_j}$, in which the operator $\Lambda:=\sum_{i=1}^kp_i\Pi_i\op{\psi_i}{\psi_i}$ is hermitian \cite{Holevo-1973a, *Yuen-1975a, *Barnett-2009a}.  Since $\sum_{i=1}^k\Pi_i=\mathbb{I}$, we have
\begin{equation*}
0=tr[\Lambda-\Lambda]=\sum_{j=1}^k tr[\Pi_j(\Lambda-p_j\rho_j)].
\end{equation*}
Then as $\Lambda-p_j\rho_j\geq 0$ and $tr[\Pi_j(\Lambda-p_j\rho_j)]=tr[\Pi^{1/2}_j(\Lambda-p_j\rho_j)\Pi^{1/2}_j]\geq 0$,
we must have that 
\begin{equation}
\label{Eq:optimal-conds2}
\Pi_j(\Lambda-p_j\op{\psi_j}{\psi_j})=(\Lambda-p_j\op{\psi_j}{\psi_j})\Pi_j=0.
\end{equation}
Our argument now proceeds analogously to the one given in Ref. \cite{Mochon-2006a}.  Let $P_S$ be the projector onto $S$, and for some POVM $\{\Pi_i\}_{i=1}^k$ that obtains $P_{opt}$ on $\mathcal{E}$, define $\hat{\Pi}_i=P_S\Pi_i P_S$.  As the $\ket{\psi_i}$ are linearly independent, there exists a set of dual states $\ket{\psi_i^\perp}$ such that $\ip{\psi_i^\perp}{\psi_j}=\delta_{ij}$.  We first note that $\Lambda-p_i\op{\psi_i}{\psi_i}\geq 0$ implies $\hat{\Lambda}-p_i\op{\psi_i}{\psi_i}\geq 0$,
where $\hat{\Lambda}=\hat{\Lambda}^\dagger=\sum_{j=1}^kp_j\hat{\Pi}_j\op{\psi_j}{\psi_j}$.  Thus, the POVM $\{\mathbb{I}-P_S,\hat{\Pi}_i\}_{i=1}^k$ also obtains $P_{opt}$ on $\mathcal{E}$.  We next note that $\hat{\Pi}_j\ket{\psi_j}\not=0$ for all $j$.  For if this were not true for some $\ket{\psi_j}$, then we could contract with $\ket{\psi^\perp_j}$ to obtain $0\leq \bra{\psi^\perp_j}  \left(\hat{\Lambda}-p_j\op{\psi_j}{\psi_j}\right)\ket{\psi^\perp_j}= -p_j$.

Next, since $\{\mathbb{I}-P_S,\hat{\Pi}_i\}_{i=1}^k$ is an optimal POVM, the corresponding equality of  Eq. \eqref{Eq:optimal-conds2} is $0=\hat{\Pi}_j(\hat{\Lambda}-p_j\op{\psi_j}{\psi_j})$.  Applying $\ket{\psi_i^\perp}$ to the RHS yields 
\begin{equation}
\hat{\Pi}_j(p_i\hat{\Pi}_i\ket{\psi_i})=\delta_{ij}p_i\hat{\Pi}_i\ket{\psi_i}.
\end{equation}
Thus, $\ket{\phi_i}:=\tfrac{1}{\sqrt{\bra{\psi_i}\hat{\Pi}_i^2\ket{\psi_i}}}\hat{\Pi}_i\ket{\psi_i}$ (which is nonzero) lies in the kernel of $\hat{\Pi}_j$ for $i\not=j$, while $\ket{\phi_i}$ is an eigenvector of $\hat{\Pi}_j$ with eigenvalue +1 when $i=j$.  Hence, $\hat{\Pi}_i=\op{\phi_i}{\phi_i}$ and $\ip{\phi_i}{\phi_j}=\delta_{ij}$, with $\sum_{i=1}^k\op{\phi_i}{\phi_i}=P_S$.  We obviously have $\bra{\phi_i}\Pi_j\ket{\phi_i}=\delta_{ij}$, which means the original POVM can perfectly distinguish the $\ket{\phi_i}$.  Conversely, any POVM $\{\Pi_i\}_{i=1}^k$ that perfectly distinguishes the $\ket{\phi_i}$ will satisfy $P_S\Pi_iP_S=\op{\phi_i}{\phi_i}$, and will therefore obtain $P_{opt}$ on $\mathcal{E}$.

Finally, let $\boldsymbol{\Pi}_S$ be the compact, convex set of POVMs with $k=dim(S)$ elements, each having support on $S$.  We have just shown that the continuous linear function $f:\boldsymbol{\Pi}_S\to\mathbb{R}$ given by $f(\{\Pi_i\}_{i=1}^k)=1-\sum_{i=1}^kp_i\bra{\psi_i}\Pi_i\ket{\psi_i}$ can be maximized only by an extreme point of $\boldsymbol{\Pi}_S$ (rank one projectors).  Convexity of $\boldsymbol{\Pi}_S$ implies that this extreme point $\mathcal{P}_0:=\{\op{\phi_i}{\phi_i}\}_{i=1}^k$ must be unique.  

(b)  For the asymptotic statement, we will need to endow $\boldsymbol{\Pi}_S$ with a metric.  For two POVMs $\mathcal{P}=\{\Pi_1,...,\Pi_k\}$ and $\mathcal{P}'=\{\Pi'_1,...,\Pi'_k\}$ in $\boldsymbol{\Pi}_S$, we can define a distance measure by $d(\mathcal{P},\mathcal{P}')=\frac{1}{2}\sum_{i=1}^k\|\Pi_i-\Pi_i'\|_1$, where $\|A\|_1={\rm Tr}|A|$ \footnote{A perhaps more natural distance measure between two POVMs is the difference in measurement probabilities, maximized over all trace one, non-negative inputs: $\max_{\rho}\frac{1}{2}\sum_{i=1}^k|tr(\rho[\Pi_i-\Pi_i'])|$.  As we are only concerned with issues of convergence, it suffices to consider the equivalent metric $d$ introduced above.}.  Note that when $\Pi_i'=\op{\phi_i'}{\phi_i'}$ is pure, we have $\frac{1}{2}\|\Pi_i-\op{\phi_i'}{\phi_i'}\|_1\geq 1-\bra{\phi_i'}\Pi_i\ket{\phi_i'}$ \cite{Nielsen-2000a}.

For any $\mathcal{P}=\{\Pi_i\}_{i=1}^k$, define $\hat{\mathcal{P}}=\{P_S\Pi_iP_S\}_{i=1}^k$.  Suppose there exists a sequence of POVMs $\mathcal{P}^{(n)}$ such that for any $\epsilon>0$, $f(\hat{\mathcal{P}}^{(n)})< P_{opt}+\epsilon$ for sufficiently large $n$.  As $\hat{\mathcal{P}}^{(n)}$ is a sequence in the compact metric space $\boldsymbol{\Pi}_S$,  by the Weierstrass Theorem from analysis, there will exist some convergent subsequence $\hat{\mathcal{P}}^{(n_j)}\to\overline{\mathcal{P}}$.  Continuity of $f$ implies that $\lim_{n_j\to \infty}f(\hat{\mathcal{P}}^{(n_j)})=f(\mathcal{P}_0)=f(\overline{\mathcal{P}})$ (recall $P_{opt}=f(\mathcal{P}_0)$).  However, by part (a), the POVM in $\boldsymbol{\Pi}_S$ obtaining $P_{opt}$ is unique and so $\overline{\mathcal{P}}=\mathcal{P}_0$.  Thus, $d(\hat{\mathcal{P}}^{(n_j)},\mathcal{P}_0)\to 0$, and so the error on $\mathcal{E}$ of each subsequence $\mathcal{P}^{(n_j)}$ satisfies
\[1-\frac{1}{k}\sum_{i=1}^k\bra{\phi_i}\Pi_i^{(n_j)}\ket{\phi_i}\leq \frac{1}{k}d(\hat{\mathcal{P}}^{(n_j)},\mathcal{P}_0)\to 0.\]  Conversely, if $1-\frac{1}{k}\sum_{i=1}^k\bra{\phi_i}\Pi_i^{(n_j)}\ket{\phi_i}\to 0$, then $1-\bra{\phi_i}\Pi_i^{(n_j)}\ket{\phi_i}\geq \tfrac{1}{4}\|\Pi_i^{(n_j)}-\op{\phi_i}{\phi_i}\|^2\to 0$ \cite{Nielsen-2000a}, which means $d(\hat{\mathcal{P}}^{(n_j)},\mathcal{P}_0)\to 0$.  By continuity of $f$, we have $1-\sum_{i=1}^kp_i\bra{\psi_i}\Pi_i^{(n_j)}\ket{\psi_i}=f(\hat{\mathcal{P}}^{(n_j)})\to P_{opt}$.

\subsection{All Conditions of Proposition \ref{Prop1} Cannot be Simultaneously Satisfied}

Condition (iii) requires orthogonality $\bra{F_i}E\ket{F_j}=0$ for $i\not=j$, and so in the basis $\{\ket{\Psi^-},\ket{F_i}\}_{i=0}^2$, $E$ must take the form
\begin{align}
\label{Eq:E}
s\op{\Psi^-}{\Psi^-}+\sum_{i=0}^2(a_i\op{F_i}{F_i}+[b_i\op{\Psi^-}{F_i}+C.C.])
\end{align}
where $s,a_i\geq 0$, $\sum_{i=0}^2a_i=1$, and $C.C.$ denotes the complex conjugate.  If $E$ is a product operator across Alice and Bob's system, then $\gamma_{01}={}_A\bra{0}E\ket{1}_{A}$ must commute with $\gamma_{10}={}_A\bra{1}E\ket{0}_{A}$.  Here we are taking partial contractions on Alice's system so that $\gamma_{01}$ and $\gamma_{10}$ are operators acting on Bob's system.  By directly computing the commutator using Eqs. \eqref{Eq:GlobalPOVM} and \eqref{Eq:E}, the condition $\bra{0}[\gamma_{01},\gamma_{10}]\ket{0}=0$ becomes
\begin{equation}
0=6[Im(b_2 -  b_3)]^2+(s+ a_0 - \tfrac{2}{3}) (s-\tfrac{1}{3}).
\end{equation}
With $a_0=\chi\geq\tfrac{1}{3}$ (condition (ii)), it is clear that $s\leq \tfrac{1}{3}$.  However, if $s<\tfrac{1}{3}$, then this equation cannot hold for any $a_0\in[\tfrac{1}{3},\tfrac{2}{3}-s)$.  Thus, the product form constraint on $E$ requires $a_0=\tfrac{1}{3}$. 

Next, we focus on the range $a_0\in[\tfrac{1}{3},\tfrac{1}{2})$, which because of (iii), guarantees that $E$ is full rank.  It is known that the $\ket{F_i'}$ can be perfectly distinguished by separable operations if and only if $\sum_{i=0}^2C(F'_i)=C(\Psi')$, where $C(\cdot)$ is the concurrence of the state and $\ket{\Psi'}=\frac{E^{-1/2} \ket{\Psi^-}}{\sqrt{\bra{\Psi^-}E^{-1} \ket{\Psi^-}}}$ (see Thm. 2 of \cite{Duan-2007a}).  We combine this with the fact that for a general two qubit state $\frac{M\otimes N\ket{\varphi}}{\sqrt{\bra{\varphi}M^\dagger M\otimes N^\dagger N\ket{\varphi}}}$, its concurrence is given by is $C(\varphi)\times\frac{|det (M)||det(N)|}{\bra{\varphi}M^\dagger M\otimes N^\dagger N\ket{\varphi}}$ \cite{Verstraete-2001a}.  Therefore after noting that $C(F_i)=1/3$ and writing $E=A\otimes B$, condition (iii) of Proposition \ref{Prop1} can be satisfied if and only if
\begin{align}
1=\frac{1}{3}\sum_i\sqrt{det(A\otimes B)}\frac{\bra{\Psi^-}A^{-1} \otimes B^{-1}\ket{\Psi^-}}{\bra{F_i} A\otimes B\ket{F_i}}.
\end{align}
To compute $\bra{\Psi^-}A^{-1} \otimes B^{-1}\ket{\Psi^-}$, we use Cramer's Rule which says $(A\otimes B)^{-1}=\frac{1}{det(A\otimes B)}Adj(A\otimes B)$, where $Adj(\cdot)$ denotes the adjugate matrix.  From \eqref{Eq:E}, we have that $\bra{\Psi^-}Adj(A\otimes B)\ket{\Psi^-}=\prod_{i=1}^3\bra{F_i} A\otimes B\ket{F_i}$.  Substituting this into the above equation gives
\begin{align}
\label{Eq:ConcurrenceConditionPostMeasure}
1=& \frac{1}{3}\sum_i\frac{1}{\sqrt{det(A\otimes B)}}\frac{\prod_{j=1}^3\bra{F_j} A\otimes B\ket{F_j}}{\bra{F_i} A\otimes B\ket{F_i}}\notag\\
\geq&\frac{1}{3}\frac{a_0a_1+a_0a_2+a_1a_2}{\sqrt{1/3a_0a_1a_2}},
\end{align}
where we have used \eqref{Eq:E} and Hadamard's inequality: $det(A\otimes B)\leq sa_0a_1a_2=1/3a_0a_1a_2$.  It is a straightforward optimization calculation to see that under the constraint $\sum_{i=0}^2a_i=1$, the RHS of \eqref{Eq:ConcurrenceConditionPostMeasure} obtains a minimum of 1 if and only if $a_0=a_1=a_2=\tfrac{1}{3}$.  This proves that condition (iii) is impossible whenever $\chi>\tfrac{1}{3}$.

\section{Eq. \eqref{Eq:mixedminerr} and the Minimum Error for one Mixed and one Pure State}

We compute an analytic formula for the minimum error probability in distinguishing weighted qubit states $\rho=p_0\op{\psi_0}{\psi_0}$ and $\sigma=p_1\op{\psi_1}{\psi_1}+p_2\op{\psi_2}{\psi_2}$.  The minimum error probability is given by $P_{err}=1/2-1/2\|\rho-\sigma\|_1$,
where $\|\cdot\|_1$ is the trace norm.  Since $\rho-\sigma$ is hermitian with eigenvalues $\lambda_i$, we have $\|\rho-\sigma\|_1=\sum_{i}|\lambda_i|$.  Thus, it is just a matter of computing the eigenvalues of $\Delta:=\rho-\sigma=p_0\op{\psi_0}{\psi_0}-p_1\op{\psi_1}{\psi_1}-p_2\op{\psi_2}{\psi_2}$.  Taking $\ket{\psi_i}=c_{i0}\ket{0}+c_{i1}\ket{1}$, we write $\Delta$ in coordinates:
\begin{align}
\Delta=&p_0\begin{pmatrix}|c_{00}|^2& c_{00}c_{01}^*\\ c_{00}^*c_{01}&|c_{01}|^2\end{pmatrix}-\sum_{i=1}^2p_i\begin{pmatrix}|c_{i0}|^2& c_{i0}c_{i1}^*\\ c_{i0}^*c_{i1}&|c_{i1}|^2\end{pmatrix}.\notag
\end{align}
For a $2\times 2$ matrix, $M=\left(\begin{smallmatrix}a&b\\c&d\end{smallmatrix}\right)$, its eigenvalues are given by the expression $\lambda_{\pm}=1/2(a+d\pm\sqrt{(a+d)^2-4\det M})$.  Thus, have that  
\[|\lambda_+|+|\lambda_-|=\begin{cases}|a+d|\;\;\;\text{if}\;\; \det M\geq 0\\\sqrt{(a+d)^2-4\det M}\;\;\;\text{if}\;\; \det M\leq 0.\end{cases}\]
Letting $M=\Delta$, we can compute that $a+d=p_0-p_1-p_2$ and 
\begin{align*}
\det \Delta=&p_1p_2(1-|\ip{\psi_1}{\psi_2}|^2)\notag\\&
-p_0p_1(1-|\ip{\psi_0}{\psi_1}|^2)-p_0p_2(1-|\ip{\psi_0}{\psi_2}|^2).
\end{align*}
Therefore, we arrive at the following
\begin{lemma}
For the weighted states $\rho$ and $\sigma$, the minimum error probability is
\begin{align}
\label{Eq:prob}
\tfrac{1}{2}-\tfrac{1}{2}\sqrt{1-4p_1p_2(1-|\ip{\psi_1}{\psi_2}|^2)-4p_0\sum_{i=1}^2p_i|\ip{\psi_0}{\psi_i}|^2}\notag
\end{align}
if $\det(\rho-\sigma)\leq 0$, and $\tfrac{1}{2}-\tfrac{1}{2}|p_0-p_1-p_2|$ if $\det(\rho-\sigma)\geq 0$.
\end{lemma}

Now, we use this result with Eq. \eqref{Eq:probsBob} to prove Eq. \eqref{Eq:mixedminerr}.  Specifically, since $|\ip{s_i}{s_j}|^2=\tfrac{1}{4}$ for the trine states, we have that $\det\Delta$ is given by
\begin{equation*}
\frac{-1}{192}(3(3+\cos(2\phi))+32\cos(2\theta)-(13+3\cos(2\phi))\cos(4\theta)).
\end{equation*}
It is straightforward to verify that this is not positive for $\phi\in [0,2\pi)$ and $\theta\in[-\pi/6,\pi/6)$.  Therefore, by Lemma 1, we obtain Eq. \eqref{Eq:mixedminerr}.

\bibliography{QuantumBib}

\end{document}